\documentclass[12pt, draftcls, onecolumn]{IEEEtran}
\usepackage{setspace}
\usepackage{multirow,booktabs}
\usepackage[dvips]{graphicx}
\usepackage[cmex10]{amsmath}
\usepackage{amsmath}
\usepackage{amssymb}
\usepackage{amsfonts}
\usepackage{fmtcount}
\DeclareGraphicsExtensions{ps,eps,pst,pdf}
\usepackage{cite}

\IEEEoverridecommandlockouts
\begin{document}

\title{Achieving Autonomous Compressive Spectrum Sensing for Cognitive Radios}

\author{Jing Jiang,~\IEEEmembership{Member,~IEEE,}
 Hongjian~Sun$^{*}$,~\IEEEmembership{Senior Member,~IEEE,}\\
 David~Baglee, and 
 H.~Vincent~Poor,~\IEEEmembership{Fellow,~IEEE}  
\vspace{-2ex}
\thanks{The paper is accepted to be published in IEEE Transactions on Vehicular Technology. This is a preprint version. Copyright (c) 2015 IEEE. Personal use of this material is permitted.  However, permission to use this material for any other purposes must be obtained from the IEEE by sending a request to pubs-permissions@ieee.org.}
\thanks{J. Jiang and D. Baglee is with the Institute for Automotive and Manufacturing Advanced Practice (AMAP), University of Sunderland, Sunderland SR5 3XB, UK. (Email: jing.jiang@sunderland.ac.uk, David.Baglee@sunderland.ac.uk)}
 \thanks{H. Sun (corresponding author) is with the School of Engineering and Computing Sciences, Durham University, Durham DH1 3LE, UK. (Email: hongjian.sun@durham.ac.uk)}
\thanks{H. V. Poor is with Department of Electrical Engineering, Princeton University, Princeton, NJ 08544, US. (Email: poor@princeton.edu)}
\thanks{The research leading to these results has received funding from the European Commision's Horizon 2020 Framework Programme (H2020/2014-2020) under grant agreement No 646470, SmarterEMC2 Project.}
}

\maketitle

\vspace{-3em}

\begin{abstract}
Compressive sensing (CS) technologies present many advantages over other existing approaches for implementing wideband spectrum sensing in cognitive radios (CRs), such as reduced sampling rate and computational complexity. However, there are two significant challenges: 1) choosing an appropriate number of sub-Nyquist measurements, and 2) deciding when to terminate the greedy recovery algorithm that reconstructs wideband spectrum. In this paper, an autonomous compressive  spectrum sensing (ACSS) framework is presented that enables a CR to automatically choose the number of measurements while guaranteeing the wideband spectrum recovery with a small predictable recovery error. This is realized by the proposed measurement infrastructure and the validation technique. The proposed ACSS can find a good spectral estimate with high confidence by using only a small testing subset in both noiseless and noisy environments. Furthermore, a sparsity-aware spectral recovery algorithm is proposed to recover the wideband spectrum without requiring knowledge of the instantaneous spectral sparsity level. Such an algorithm bridges the gap between CS theory and practical spectrum sensing. Simulation results show that ACSS can not only recover the spectrum using an appropriate number of measurements, but can also considerably improve the spectral recovery performance compared with existing CS approaches. The proposed recovery algorithm can autonomously adopt a proper number of iterations, therefore solving the problems of under-fitting or over-fitting which commonly exist in most greedy recovery algorithms.
\end{abstract}


\section{Introduction}
\label{section1}

The radio frequency (RF) spectrum is a finite natural resource, currently regulated by government agencies. According to current policy, primary user (PU) on a particular spectrum band has exclusive right to the licensed spectrum. With the explosive growth of wireless applications, the demands for RF spectrum are constantly increasing. On the other hand, it has been reported that localized temporal and geographic spectrum utilization efficiency is extremely low~\cite{nsf, surveyme}. Cognitive radio (CR) \cite{bookcr} has emerged as one of the most promising solutions that address the spectral under-utilization problem. A crucial requirement of CRs is that they must rapidly exploit spectrum holes (i.e., portions of the licensed spectrum that are not being used by PUs) without causing harmful interference to PUs. This task is achieved by spectrum sensing, which can be defined as a technique for achieving awareness about the spectral usage and existence of PUs in a given geographical area \cite{sunTSP, chapter6}.

CR with a wide spectral awareness (e.g., a few GHz rather than MHz) could potentially exploit more spectral opportunities and achieve larger capacity. Wideband spectrum sensing techniques (categorized into Nyquist wideband sensing and sub-Nyquist wideband sensing) therefore have attracted considerable attention in research on CR networks\cite{surveyme}. In~\cite{Tian2006}, Tian and Giannakis proposed a wavelet based approach using Nyquist sampling rate for wideband spectrum sensing. Quan {\em et al.} \cite{quan2, quan} presented a multiband joint detection (MJD) approach to detect the primary signal from Nyquist samples over multiple frequency bands. 
Note that according to the Nyquist sampling theory, the received signal at CR should be sampled at a sampling rate of at least twice the maximum signal frequency \cite{sunTSP}. Thus, to achieve a ``wider'' spectral awareness at CRs (i.e., a larger signal frequency range), a high sampling rate is needed, leading to excessive memory requirements and high energy cost. This motivates the development of sub-Nyquist technologies (using sampling rates lower than the Nyquist rate) for reducing the operational sampling rate while retaining the spectral information\cite{cs, RIP}. 

The compressive sensing (CS) theory was first introduced to implement the sub-Nyquist spectrum sensing in CR networks in \cite{scs1}. This technique used a number of samples closer to the information rate and reconstructed the wideband spectrum using these partial measurements. 
Note that using CS techniques, the wideband signal to be sampled is required to be sparse in a suitable basis \cite{CS1, CS2}; this requirement can typically be met in CR networks due to the low spectral occupancy\cite{surveyme}. Several sub-Nyquist wideband spectrum sensing algorithms were proposed to mitigate the effects of multipath fading in cooperative CR networks in \cite{wide3, me, zeng,  wide8}.  After sub-Nyquist sampling, the wideband signal can be recovered from these sub-Nyquist samples by using one of several possible recovery algorithms, e.g., orthogonal matching pursuit (OMP) \cite{Tropp2007, omp2} or compressive sampling matching pursuit (CoSaMP)\cite{cosamp, Cosamp1}. Given a known sparsity level such as $k$, an appropriate number of measurements (samples)  $M=C_0 k\log (N/k)$ can be chosen such that  the quality of recovery can be secured, where $C_0$ denotes a constant and $N$ denotes the number of measurements if the Nyquist rate is utilized.
Consequently, such CS-based algorithms can take advantage of using sub-Nyquist sampling rates for signal acquisition, instead of the Nyquist rate, leading to reduced energy consumption, complexity, and memory requirements. 

\begin{figure}[ht]
\centerline{\includegraphics[width=5.6in]{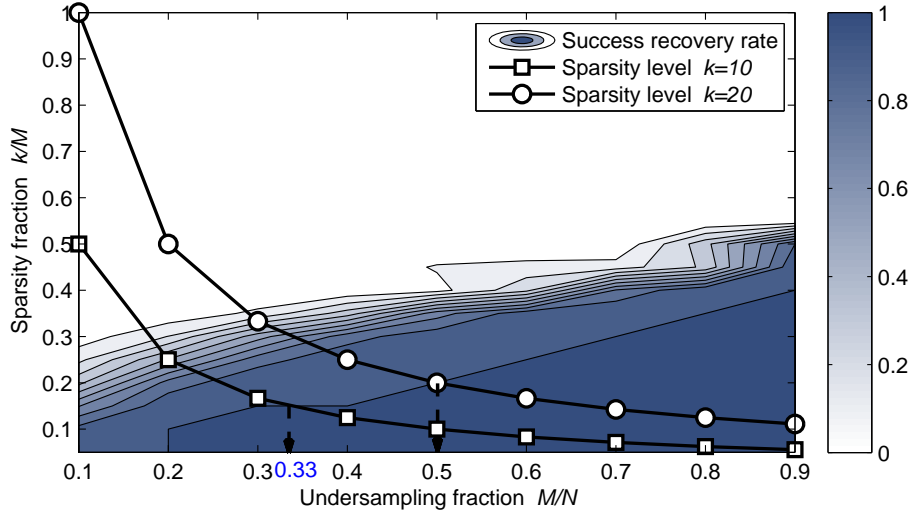}}
\vspace{-1em}
\caption{In a traditional CS system, the successful recovery rate varies when the number of measurements and the sparsity level vary \cite{chapter6}. In simulations, we assumed $N=200$ and varied the number of measurements $M$ from 20 to 180 in eight equal-length steps. Additionally, we chose the sparsity level $k\in[1,M]$ and adopted Gaussian measurement matrices. After 5000 trials of each parameter setting, we obtained this figure. }
\label{fig1}
\end{figure}

It is worthwhile to emphasize that directly applying CS theory to CR networks may lose its inherent advantages in practice. This is because to guarantee a high successful recovery rate, CS approaches tend to pessimistically choose the number of measurements $M$ larger than that is necessary: For example, as depicted in Fig.~\ref{fig1}, when $k=10$, $M=33\%N$ can be used for guaranteeing a very high successful recovery rate; but this is not always necessary because by using fewer measurements we may still recover the spectrum with an appropriate or predefined probability. 
Most importantly, the number of measurements $M$ is always linked to the spectrum sparsity level $k$, which means the knowledge of $k$ will be required for determining an appropriate value of $M$ in CR networks. However, the sparsity level of the radio spectrum is often unknown due to either the dynamic activities of PUs or the time-varying fading channels between PUs and CRs\cite{surveyme}.
Because of this sparsity level uncertainty in practical CR networks, most CS approaches intend to further increase measurements to ensure a high successful recovery rate, thereby leading to more unnecessary energy consumption. For example, in Fig.~\ref{fig1}, for the uncertainty range $10 \le k \le 20$, $M=50\%N$ (rather than $M=33\%N$) will be selected, which does not fully exploit the inherent advantages of using CS techniques for implementing wideband spectrum sensing in CR networks.

Against the aforementioned background, this paper aims to bridge the gap between CS theory and practical spectrum sensing. In particular, the novel contributions of this paper can be summarized as follows:
\begin{itemize}
\item An autonomous compressive  spectrum sensing (ACSS) framework is proposed for recovering the wideband spectrum by using an appropriate number of compressive measurements. This framework does not require prior knowledge of the instantaneous spectral sparsity level, resulting in reduced system complexity. Performance analysis is given to show that the proposed ACSS framework can inherently avoid excessive or insufficient numbers of compressive measurements, and help improve CR system throughputs.
\item A novel validation approach is proposed to accurately estimate the actual spectral recovery error with high confidence by using only a small amount of testing data. Note that the actual spectral recovery error is typically unknown as the actual wideband spectrum is not accessible under sub-Nyquist rate. This validation approach applied in the ACSS framework enables compressive measurement acquisition halted at an earliest appropriate time\footnote[1]{Please note that Bayesian compressive sensing \cite{bcs, Bishop2007} can also simultaneously perform reconstruction and validation, and determine the confidence level of estimation results.}. 
\item  To extend the use of ACSS to noisy measurement environments, another validation method is proposed. Theoretical analysis shows that, if a good spectral estimate exists, the proposed validation method can find it with a very high probability by using a small testing subset. 
\item A sparsity-aware spectral recovery algorithm is designed for spectral recovery without  requiring knowledge of the instantaneous spectral sparsity level. Iterations of the recovery algorithm are analyzed and shown to be able to terminate at the correct iteration index, which therefore reduces the possibilities of under-/over-fitting. 
\end{itemize}

The rest of the paper is organized as follows. Section~\ref{section2} introduces compressive spectrum sensing problems and the system model. Section \ref{section3} presents the ACSS framework and analyzes its halting criterion. ACSS is then applied and analyzed in noisy environments in Section~\ref{section4}, and the sparsity-aware recovery algorithm is proposed in Section \ref{section4.2}. Simulation results are presented in Section \ref{section5}, with conclusions in Section \ref{section6}. We note that, throughout this paper, letters with horizontal arrows above them are used to represent vectors, e.g., $\vec{x}$ and $\vec{X}$ where the lowercase letter denotes the time-domain and the uppercase letter denotes the Fourier domain. Uppercase boldface letters are used to denote matrices, e.g., $\mathbf{\Phi}$. And an $N \times N$ discrete Fourier transform (DFT) matrix is denoted by $\mathbf{F}_N$, where $\mathbf{F}_N^{-1}$ denotes the inverse of the matrix $\mathbf{F}_N$.

\section{System Model and Problem Statement}
\label{section2}

Consider that a CR node receives an analog signal $x(t)$ from PUs, which has the frequency range $0 \sim W$ Hz. 
Based on the Nyquist sampling theory, such an analog signal should be sampled at the sampling rate $f \ge 2W$ Hz. After a small time step~$\tau$ (seconds) of Nyquist sampling, we will obtain a full signal vector $\vec{x}\in \mathbb{C}^{N\times 1}$, where $N=f\tau$ (an integer number by properly choosing the sampling rate) denotes the number of samples.

CS theory indicates that a sparse signal can be acquired by using a sub-Nyquist sampling rate $f_s$ ($f_s<2W$), which results in fewer samples than predicted on the basis of Nyquist sampling theory. 
The value of $f_s$ is determined by the potential under-sampling fraction multiplying $f$. 
Since the spectrum is often sparse in CR networks due to the low spectral occupancy \cite{scs1}, CS theory has been applied for signal acquisition at CRs \cite{wide3, wide6, me}.
Here, the use of a sub-Nyquist sampler, such as the random demodulator \cite{beyond}, will generate a compressive measurement vector $\vec{y}\in \mathbb{C}^{M\times 1}$ ($M=f_s\tau < N$). Mathematically, the compressive measurement vector $\vec{y}$ can be written as $\vec{y}=\mathbf{\Phi} \vec{x}$, where $\vec{x}$ denotes the signal vector if the Nyquist rate is employed, and $\mathbf{\Phi}$ denotes an $M\times N$ measurement matrix that can be implemented using a sub-Nyquist sampler.  If the signal $\vec{x}$ is $k$-sparse ($k <M< N$) in some basis and the measurement matrix is appropriate, we can recover $\vec{x}$ from $\vec{y}$ using recovery algorithms. This actually means that, using CS theory, we can obtain $\vec{x}$ by merely using the sub-Nyquist sampling rate $f_s$, instead of the Nyquist sampling rate $f$. 

\begin{figure}[ht]
\centering
\centerline{\includegraphics[width=5.5 in]{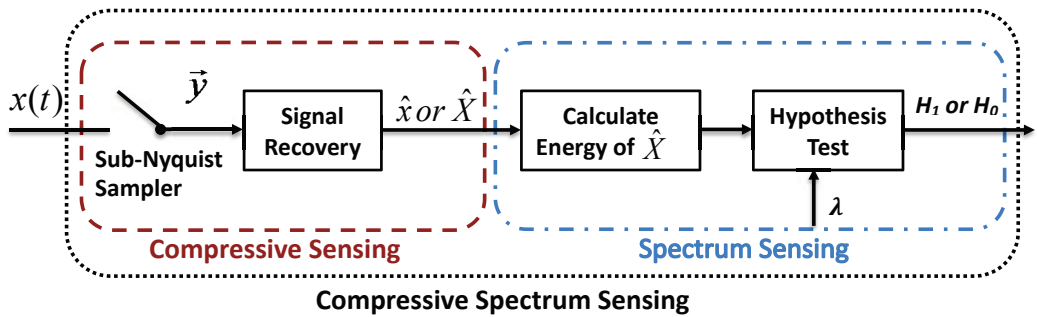}}
\vspace{-0.8em}
\caption{Diagram of compressive spectrum sensing: The spectral domain energy detection approach is used for spectrum sensing.}
\label{fig2}
\end{figure}

The basic structure of CS-based spectrum sensing (also called compressive spectrum sensing) used in this paper is shown in Fig.~\ref{fig2}. The aim is to recover $\vec{x}$ and its DFT spectrum $\vec{X}=\mathbf{F}_N\vec{x}$ from compressive measurements $\vec{y}$, and then perform spectrum sensing using the recovered signal $\hat{x}$ or its DFT spectrum $\hat{X}$. For an overview of state-of-the-art compressive spectrum sensing techniques, the reader is referred to \cite{surveyme}. 
Spectral domain energy detection \cite{hongjian} is a typical spectrum sensing approach, and thus is adopted in this paper. As shown in Fig.~\ref{fig2}, using this approach, we can extract the recovered spectrum within the frequency range of interest (e.g., $\Delta f$) and calculate its signal energy.  A detection threshold (denoted by $\lambda$) is then chosen and compared with the signal energy to decide whether this frequency band is occupied or not, i.e., choosing between binary hypotheses $\mathcal{H}_{1}$ (occupied) and $\mathcal{H}_{0}$ (not occupied).

According to the structure of compressive spectrum sensing, we know that the recovery quality will have significant impact on the performance of compressive spectrum sensing. The recovery quality depends on the following factors: the sparsity level, the choice of measurement matrix, the recovery algorithm, and the number of compressive measurements. The sparsity level of spectrum in CR networks is mainly determined by the PUs' activities within a frequency range and the medium access control (MAC) of the CRs. To evaluate the suitability of a chosen measurement matrix, we adopt an elegant metric: the restricted isometry property (RIP) \cite{RIP}. In \cite{beyond} and  \cite{was1}, sub-Nyquist samplers with controllable measurement matrices have been proposed to realize CS. Using such samplers, the primary signal received at CRs is first modulated by pseudo-random sequences (which are determined by pseudo-random seeds), and then sampled by standard low-rate samplers. Since these pseudo-random sequences are known and controllable, we can easily construct known measurement matrices subject to satisfactory RIP. For a comprehensive understanding of RIP and measurement matrix design, the reader is referred to \cite{rip2, rip3} and \cite{mad1, mad2}, respectively. In the rest of this paper, we will thus focus on discussing the following two factors: the number of measurements and the recovery algorithm.

\section{Autonomous Compressive  Spectrum Sensing (ACSS)}
\label{section3}

In this section, we will propose the ACSS framework enabling us to gradually acquire compressive measurements using the sub-Nyquist sampling rate, recover the DFT spectrum, and halt the compressive measurements at the correct time. The halting criterion and performance analysis will be provided to show that ACSS can avoid excessive or insufficient numbers of compressive measurements.

\subsection{Model and Framework of ACSS}
\label{section3.1}

\begin{figure}[ht]
\centering
\centerline{\includegraphics[width=4.6in]{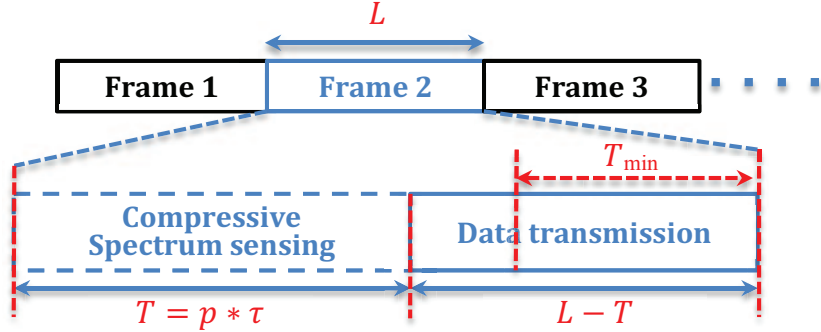}}
\vspace{-0.7em}
\caption{Frame infrastructure of periodic spectrum sensing in cognitive radio networks.}
\label{fig3}
\end{figure}

Consider that CR networks utilize a periodic spectrum sensing structure and each time frame has a fixed length $L$ (seconds) which consists of a spectrum sensing time slot and a data transmission time slot, as depicted in Fig.~\ref{fig3}. The spectrum sensing duration $T$ ($0<T<L$) is adjustable and equals $p$ (a positive integer) times as long as the small time step $\tau$, i.e., $T=p\tau$. To guarantee the bit rate at CRs, at least $T_{min}$ (seconds) should be reserved for data transmission; thus, the spectrum sensing duration $T$ will satisfy $L-T=L-p\tau \ge T_{min}$, equivalently, $p\le \frac{L-T_{min}}{\tau}$. Here, we assume that the spectrum sensing duration $T$ is smaller than the channel coherence time, such that the magnitude of the channel response remains constant within $T$. In addition, we assume that, within $T$, the primary signals are wide-sense stationarity and all CRs can keep quiet as enforced by protocols (e.g., at the MAC layer \cite{quan2}). This means that the spectral components of the DFT spectrum $\vec{X}=\mathbf{F}_N\vec{x}$ arise only from PUs and background noise. Due to the low spectral occupancy in CR networks~\cite{scs1}, the DFT spectrum $\vec{X}$ can be assumed to be $k$-sparse, which means the spectrum consists only of the $k$ largest values that cannot be ignored.
This sparsity level $k$ is typically unknown but has a known upper bound $k_{\max}$. This is because, in practice, the instantaneous spectral occupancy may be difficult to obtain, but the maximal spectral occupancy can be easily estimated by long-term spectral usage measurements. For example, the maximal spectral occupancy within 30~MHz - 3~GHz in New York City has been reported to be only $13.1\%$~\cite{nsf}. In such a scenario, $k_{\max}$ can be calculated by $k_{\max}=13.1\%\times N$.

Using ACSS, we perform compressive measurements using the sub-Nyquist sampling rate $f_s$ ($f_s<2W$). The same sub-Nyquist sampler is adopted throughout the spectrum sensing duration $T$, and the corresponding measurement matrices follow the same distribution, e.g., the standard normal distribution, or the Bernoulli distribution\footnote[2]{It has been proved in \cite{cs} and \cite{RIP} that, if the number of measurements is appropriate, the measurement matrix with either Gaussian or Bernoulli distribution can secure the RIP condition with an overwhelming probability.}  with equal probability on $\pm 1$\cite{cs, RIP}. Furthermore, the set of compressive samples within $T$ is denoted by $\vec{y}_{p}$ ($\vec{y}_{p} \in \mathbb{C}^{M_{p}\times 1}$), where $M_{p}=f_s T=f_s p \tau$ is the number of compressive measurements. 
The set of compressive samples $\vec{y}_{p}$ is then divided into two subsets including the training subset $\vec{R}_{p}$ ($\vec{R}_{p} \in \mathbb{C}^{r_{p}\times 1}$) to recover the  spectrum, and the testing subset $\vec{V}_{p}$ ($\vec{V}_{p} \in \mathbb{C}^{v_{p}\times 1}$) to validate the recovered spectrum, where $M_{p}=r_{p}+v_{p}$ and there is a trade-off\footnote[3]{Given a fixed value of $M_p$, a larger value of $v_p$ could result in higher probability of finding the best spectral approximation; while on the other hand, it leads to worse spectral recovery since $r_p=M_p - v_p$ becomes less.}  between $v_p$ and $r_p$. 
Based on CS theory, the two subsets can be expressed as
\begin{equation}
\vec{R}_{p} =\mathbf{\Phi}_{p} \vec{x}_{p}=\mathbf{\Phi}_{p} \mathbf{F}_{pN}^{-1} \vec{X}_{p} ,
\label{train}
\end{equation}
and
\begin{equation}
\vec{V}_{p} =\mathbf{\Psi}_{p} \vec{x}_{p}=\mathbf{\Psi}_{p} \mathbf{F}_{pN}^{-1} \vec{X}_{p} ,
\label{test}
\end{equation}
respectively, where $\mathbf{\Phi}_{p}$ is an $r_{p}\times pN$ measurement matrix, $\vec{x}_{p}\in \mathbb{C}^{pN\times 1}$ denotes the signal vector if the Nyquist sampling rate is used within $T$, $\vec{X}_{p}$ denotes the DFT spectrum of $\vec{x}_{p}$ such that $\vec{X}_{p}=\mathbf{F}_{pN}\vec{x}_{p}$, and $\mathbf{\Psi}_{p}$ is a $v_{p}\times pN$ testing matrix.
Using the OMP recovery algorithm in \cite{Tropp2007, omp2}, we could obtain a spectral estimate $\hat{X}_{p}$ from $\vec{R}_{p}$. When we adjust the spectrum sensing duration $T=p\tau$ step by step (via increasing $p$), a sequence of spectral estimates, i.e., $\hat{X}_{1}, \hat{X}_{2}, \cdots, \hat{X}_{p}$, will be obtained. The compressive sampling will be halted once a satisfactory spectral estimate is found that meets the halting criterion, or the satisfactory spectral estimate cannot be found within the given time.

The work flow of ACSS is shown in Table~\ref{table:a1}. The halting criterion will be analyzed in Section \ref{section3.2}. We emphasize that unlike traditional CS approaches, 
the proposed ACSS divides the spectrum sensing duration into several mini time slots, performs compressive sampling step by step, and halts the sampling at an earliest appropriate time (once an appropriate spectral estimate is found). In this case, some spectrum sensing time slots can be saved and then used for data transmission, which will not only improve the CR system throughput (by using longer transmission time) but also save energy used for spectrum sensing. Furthermore, unlike other CS approaches, the proposed ACSS does not require the knowledge of the spectral sparsity level because of the introduction of a validation procedure, where the compressive samples obtained during one time step are divided into two subsets and a small testing subset is used for validation. The proposed halting criterion enables the sampling to be terminated at the earliest appropriate time while guaranteeing wideband spectrum recovery with a small predictable recovery error.

\begin{table}[!t]
\centering
{\small
\tabcolsep 3pt
\caption{Work Flow of The ACSS Framework} 
\begin{tabular}{|l|} 
\hline\hline 
\textbf{Inputs}\\\hspace{0.8em} 
Frame length $L$, minimum data transmission duration \\
\hspace{0.8em} $T_{min}$, sampling rate $f_s$, time step $\tau$, size of testing\\
\hspace{0.8em} measurements $v_p$, recovery error threshold $\varpi$, \\
\hspace{0.8em} confidence factor $\eta$, energy detection threshold $\lambda$.\\ \hline
1. \textbf{Initialize}  the time step index $p=1$.\\
2. \textbf{Repeat} 
\\
\hspace{0.8em} a). perform compressive sampling using $f_s$, obtaining \\
\hspace{2.2em}  the measurement set $\vec{y}_p$; \\
\hspace{0.8em} b). partition $\vec{y}_p$ into the training subset $\vec{R}_p$ and the \\\hspace{2.2em}  testing subset $\vec{V}_p$ $^4$; \\
\hspace{0.8em} c). use a spectral recovery algorithm to estimate the\\
\hspace{2.2em}  spectrum from $\vec{R}_p$, and obtain the spectral estimate \\
\hspace{2.2em} $\hat{X}_p$; \\
\hspace{0.8em} d). calculate and update the validation parameter \\
\hspace{2.2em} $\rho_p = \frac{\|\vec{V}_p-\mathbf{\Psi}_p \mathbf{F}_{pN}^{-1} \hat{X}_p \|_1}{v_p}$;\\
\hspace{0.8em} e). update the time step index $p=p+1$. \\
3. \textbf{Until} the halting criterion $ \rho_p \le \varpi ({1-\eta}) \sqrt{\frac{2}{\pi pN}} $ is \\
\hspace{0.8em} true, or $p > \frac{L-T_{min}}{\tau}$. \\
4. \textbf{Stop} sub-Nyquist compressive sampling.\\\hline
5. \textbf{If} the halting criterion is true, \\
\hspace{0.8em} a) perform energy detection $\|\hat{X}_p\|^2 \underset{\mathcal{H}_0}{\overset{\mathcal{H}_1}{\gtrless}} \lambda $;  \\
\hspace{0.8em} b) \textbf{for} ${\mathcal{H}_0}$, transmit data via un-occupied bands. \\
\hspace{2em} \textbf{for} ${\mathcal{H}_1}$, return and report the spectrum is occupied. \\
\hspace{1em} \textbf{Else} \\
\hspace{1em} Increase $f_s$ and wait for next spectrum sensing frame.\\
\hspace{1em} \textbf{End}\\
\hline \hline  
\end{tabular}
\label{table:a1}  
}
\end{table}
\addtocounter{footnote}{4}
\footnotetext[\value{footnote}]{The size of the testing subset $v_p$ is given as an input, which is chosen according to the following Lemma 1 in the noiseless case or Theorem 2 in the noisy case. We then have the size of the training subset $r_p=M_p-v_p$.}

\subsection{Halting Criterion and Performance Analysis}
\label{section3.2}

As shown in Table~\ref{table:a1}, the halting criterion plays a crucial role in determining the performance of the ACSS framework. To improve the energy efficiency of CRs, we hope that the compressive sampling can be halted at the earliest appropriate time such that the current spectral estimate $\hat{X}_p$ is a good estimate to $\vec{X}_p$ (i.e., the spectral recovery error $\|\vec{X}_p-\hat{X}_p \|_2$ is sufficiently small). 
However, the spectral recovery error $\|\vec{X}_p-\hat{X}_p \|_2$ is typically not known because the real DFT $\vec{X}_p$ is unknown under the sub-Nyquist sampling rate. Thus, using traditional CS approaches, we do not know when we should halt the compressive sampling. To solve this problem, we define the validation parameter $\rho_p$ to serve as a proxy for the actual recovery error:
\begin{equation}
\rho_p \stackrel{\triangle}{=} \frac{\| \vec{V}_p-\mathbf{\Psi}_p \mathbf{F}_{pN}^{-1} \hat{X}_p \|_1}{v_p},
\label{valid}
\end{equation}
In the following lemma, we give a result on the relationship between the validation parameter $\rho_p$ and the actual spectral recovery error $\|\vec{X}_p-\hat{X}_p \|_2$:

{\bf Lemma~1}\footnote[5]{In CS, an estimate $\hat{x}$ can be obtained by using an $\ell_1$ or mixed $\ell_1/\ell_2$-based recovery algorithm. However, the similarity/difference between $\hat{x}$ and the actual signal $\vec{x}$ is not known because the actual signal cannot be directly obtained under the sub-Nyquist rate. This lemma aims to find how far $\hat{x}$ is from $\vec{x}$ (equivalently $\hat{X}$ from $\vec{X}$ ) by considering the $\ell_2$ metric $\|\hat{x}-\vec{x}\|_2$, in order to halt compressive sampling for saving energy at CRs.}:  
For a given confidence factor $\eta\in(0, \frac{1}{2})$, $\xi \in (0, 1)$, $v_p=C\eta^{-2}\log \frac{4}{\xi}$ where $C$ denotes a positive constant, the confidence interval $\left[\frac{\sqrt{\frac{\pi pN}{2}}\rho_p}{1+\eta}, \frac{\sqrt{\frac{\pi pN}{2}}\rho_p}{1-\eta}\right]$ can act as a good estimate of the unknown parameter $\|\vec{X}_p-\hat{X}_p \|_2$ such that 
\begin{equation} 
\Pr \left[\frac{\sqrt{\frac{\pi pN}{2}}\rho_p}{1+\eta} \le \|\vec{X}_p-\hat{X}_p\|_2 \le \frac{\sqrt{\frac{\pi pN}{2}}\rho_p}{1-\eta} \right] \ge 1-\xi , \label{the1}
\end{equation}
where the minimum confidence level $1-\xi$ can also be written as $1-4 \exp(-\frac{v_p\eta^2}{C})$ when $v_p$ is given.

See Appendix A for the proof of Lemma~1.

\emph{Remark \ref{section3}.1:} We see that the actual spectral recovery error $\|\vec{X}_p-\hat{X}_p \|_2$ can be directly linked to the validation parameter $\rho_p$ in (\ref{the1}). Even though the actual spectral recovery error $\|\vec{X}_p-\hat{X}_p \|_2$ is not known, we can predict that it lies in a known confidence interval $\left[ \frac{\sqrt{\frac{\pi pN}{2}}\rho_p}{1+\eta}, \frac{\sqrt{\frac{\pi pN}{2}}\rho_p}{1-\eta}\right]$ with a confidence level higher than $1-4 \exp(-\frac{v_p\eta^2}{C})$.
The confidence factor $\eta$ determines the width of the confidence interval how uncertain we know about the unknown spectral recovery. For a given $\eta$, increasing the value of $v_p$ (i.e., using more measurements for validation) will help to improve the confidence level.
Additionally, we note that the choice of the parameter $C$ depends on the concentration property of random variables in the matrix $\mathbf{\Psi}$\cite{Baraniuk2008}. Given a good $\mathbf{\Psi}$, e.g., the testing matrix with random variables following either the Gaussian or Bernoulli distribution as used in this paper, $C$ can be a small positive constant. The benefit of the proposed algorithm will change with different testing matrices: This is because, given the confidence factor $\eta$ and the size of the testing set $v_p$, different testing matrices will lead to different values of $C$, and thus result in different confidence levels.

{\bf Theorem~1}: Using the proposed ACSS, for a given confidence factor $\eta\in(0, \frac{1}{2})$ and spectral recovery error threshold $\varpi$, if the halting criterion $\rho_p \le \varpi ({1-\eta}) \sqrt{\frac{2}{\pi pN}}$ is met, we can find a good spectral estimate such that $\|\vec{X}_p-\hat{X}_p \|_2 \le \varpi$ with a probability higher than $1-4 \exp(-\frac{v_p\eta^2}{C})$.

See Appendix B for the proof of Theorem~1.

\emph{Remark \ref{section3}.2:} 
We can see that, using ACSS, the probability of finding a good spectral estimate exponentially grows as $v_p$ increases, i.e., as more compressive measurements are used for validation. Once the halting criterion has been met, the compressive sampling will be immediately halted as shown in Table~\ref{table:a1}. 
Furthermore, we note that Theorem 1 can be reshaped when the minimum confidence level is given. That is, to find a good spectral estimate such that $\|\vec{X}_p-\hat{X}_p \|_2 \le \varpi$ with a confidence level higher than $1-\xi$, we use the halting criterion 
\begin{equation} 
\rho_p \le \varpi\left( 1-\sqrt{\frac{C}{v_p}\log{\frac{4}{\xi}}}\, \right)  \sqrt{\frac{2}{\pi pN}}\;.
\end{equation}
From the relationship between the halting criterion and the ACSS performance as given in Theorem 1, we can see that this ACSS framework can decrease the probabilities of excessive or insufficient numbers of compressive measurements.

\section{ACSS in Noisy Environments}
\label{section4}

When performing compressive spectrum sensing, there may exist measurement noise due to the quantization error of analog-to-digital converters or the imperfect design of sub-Nyquist samplers. 
In this section, we extend the use of ACSS to such noisy environments, and will analyze the validation approach to fit the proposed framework.

Given the noisy compressive measurements, the training subset $\vec{R}_p$ and the testing subset $\vec{V}_p$ can be written as
\begin{equation}
\vec{R}_p= \mathbf{\Phi}_p \mathbf{F}_{pN}^{-1} \vec{X}_p+ \vec{n}_R ,
\label{train1}
\end{equation}
and
\begin{equation}
\vec{V}_p =\mathbf{\Psi}_p \mathbf{F}_{pN}^{-1} \vec{X}_p+\vec{n}_V ,
\label{test1}
\end{equation}
respectively,  where $\vec{n}_R$ and $\vec{n}_V$ denote the measurement noise introduced during the compressive measurement (e.g. generated by signal quantization). Without loss of generality, we model both $\vec{n}_R$ and $\vec{n}_V$ as circular complex additive white Gaussian noise (AWGN) with their components obeying a distribution $\mathcal{CN}(0, \delta^2)$.

We expect that compressive sampling can be halted if the current spectral estimate $\hat{X}_p$ is very close to the actual spectrum $\vec{X}_p$. To find this good spectral estimate, we adopt the halting criterion $|\rho_p-\sqrt{\frac{\pi}{2}}\delta|\le \theta$  due to the following: 

{\bf Theorem~2:}
Using ACSS in noisy environments, for any accuracy parameter $\theta>0$, $\delta>0$, $\varrho \in (0,1)$, 
and $v_p=\ln \left(\frac{2}{\varrho}\right) \frac{(4-\pi)\delta^2+2 \theta \delta }{\theta^2}$, to find a good spectral estimate such that $\hat{X}_p$ is sufficiently close to the actual spectrum $\vec{X}_p$, the halting criterion satisfies
\begin{equation}
\Pr \left[ \vert\rho_p-\sqrt{\frac{\pi}{2}}\delta\vert\le \theta \right] > 1-\varrho, \label{t2}
\end{equation}
where the minimum probability $1-\varrho$ can also be written as $1-\varrho=1- 2\exp \left(-\frac{v_p\theta^2}{(4-\pi)\delta^2+2  \theta \delta } \right)$.

The proof of Theorem~2 is given in Appendix C.

\emph{Remark \ref{section4}.1:} 
Theorem~2 addresses the issue of finding a good approximation of $\vec{X}_p$ in the noisy case by using the halting criterion $|\rho_p-\sqrt{\frac{\pi}{2}}\delta|\le \theta$. The accuracy parameter $\theta$ in Theorem 2 has a known relationship with the parameters $v_p$, $\delta$, and $\varrho$. 
Given a fixed confidence level $1-\varrho$, there is a trade-off between $\theta$ and the size of the testing set $v_p$: at the expense of accuracy (i.e., a large value of $\theta$), $v_p$ can be small. Additionally, we find that the probability of $|\rho_p-\sqrt{\frac{\pi}{2}}\delta|\le \theta$ rapidly increases as $v_p$ increases. That is, using more measurements for validation, we have a higher probability of finding the good spectral estimate.
   
Taking advantage of Theorem 2, we extend the use of ACSS (based on Table I) to noisy environments. The inputs in Table I will be adjusted to `\textit{frame length $L$, minimum data transmission duration $T_{min}$,  sampling rate $f_s$, time step $\tau$, size of testing measurements $v_p$, accuracy parameter $\theta$, noise variance $\delta$,  and energy detection threshold $\lambda$.}' The whole work flow of ACSS in noisy environments remains the same as in Table I except that the halting criterion is changed to $|\rho_p-\sqrt{\frac{\pi}{2}}\delta|\le \theta$. Using the proposed ACSS under the condition that the spectral sparsity level is unknown and the effects of measurements noise are not negligible,  compressive sampling can still be halted in the correct time and  the problems of excessive or insufficient numbers of measurements can be avoided.

\section{Sparsity-Aware Spectral Recovery (SASR) Algorithm}
\label{section4.2}

Traditionally, greedy recovery algorithms, e.g., OMP, will iteratively generate a sequence of estimates $\hat{X}_p^1, \hat{X}_p^2, \cdots, \hat{X}_p^t$ which can lead to a good spectral estimate under certain system parameter choices. Using $t=k$ iterations in OMP, we can obtain a $k$-sparse vector $\hat{X}_p^k$ as an estimate of the actual spectrum $\vec{X}_p$\cite{Tropp2007}. That is, the sparsity level $k$ is required to be an input for OMP, and this input is usually required in most other greedy recovery algorithms. However, in CR systems, the spectral sparsity level $k$ is often unknown or difficult to estimate, which  can result in early or late termination of that traditional greedy algorithms (i.e. underfitting and overfitting problems). On the other hand, we note that the proposed Theorem 1 and Theorem 2 are used to identify a satisfactory spectral approximation of the actual spectrum from an estimate sequence by using appropriate halting criteria. The theorems thus can be applied in recovery algorithms to solve the underfitting and overfitting problems: The halting criteria can help terminate the iterations at an appropriate time without requiring the knowledge of $k$, and an estimate of the spectrum (i.e. the recovered spectrum) will be obtained. To this end, we propose a so-called sparsity-aware spectral recovery (SASR) algorithm to handle the spectrum recovery problem given unknown instantaneous spectral sparsity level $k$, as shown in Table~\ref{table:a3}. 

Using recovery algorithms, we aim to obtain an estimate of $\vec{x}_{p}$ or its spectrum $\vec{X}_p$ from $\vec{R}_p$. Since $\vec{x}_{p}$ is $k$-sparse (i.e. $\vec{x}_{p}$ has $k$ non-zero components), the vector $\vec{R}_p=\mathbf{\Phi}_{p} \vec{x}_{p}$ is a linear combination of $k$ columns from $\mathbf{\Phi}_{p}$. We thus need to identify which column of $\mathbf{\Phi}_{p}$ is involved in $\vec{R}_p$, by choosing the column of $\mathbf{\Phi}_{p}$ that is mostly correlated to the residual of $\vec{R}_p$ at each iteration. As shown in Table~\ref{table:a3}, using the proposed SASR algorithm, we find the support index $\varphi ^t$ that can maximize the correlation between the remaining part of $\vec{R}_p$ and the measurement matrix at each iteration. A new support index set $\Omega ^t$ is then formed by merging the previously computed support index set with the current support index $\varphi ^t$.  
In the step 2-$d$) of Table~\ref{table:a3}, we note that $\mathbf{\Phi}_{p}(\Omega ^t)$ denotes a sub-matrix of  $\mathbf{\Phi}_p$ that is obtained by selecting only those columns whose indices are within $\Omega ^t$ and setting the remaining columns to zeros. We use the Moore-Penrose pseudoinverse to solve the least squares problem, and then obtain a new spectral estimate $\hat{X}_p^t$.
To verify whether $\hat{X}_p^t$ is a good spectral estimate, we calculate the parameter $\rho_p^t$ using the testing subset $\vec{V}_p$ and the spectral estimate $\hat{X}_p^t$. After that, the residual  $\vec{\gamma}_p^t$ is updated and the algorithm iterates on the residual. Finally, the spectral estimate that breaks the loop of step 2 is returned as the output of SASR algorithm.

\begin{table}[!t]
\centering
{\small
\tabcolsep 3pt
\caption{Sparsity-Aware Spectral Recovery (SASR) Algorithm} 
\begin{tabular}{|l|}  
\hline\hline 
\textbf{Inputs}: \\\hspace{0.8em} Training subset $\vec{R}_p$, testing subset $\vec{V}_p$, testing matrix   \\\hspace{0.8em} $\mathbf{\Psi}_p$, measurement matrix $\mathbf{\Phi}_p$, recovery error threshold \\
\hspace{0.8em} $\varpi$ (noiseless case), confidence factor $\eta$ (noiseless case),  \\\hspace{0.8em}  noise variance $\delta^2$ (noisy case), accuracy parameter $\theta$ \\\hspace{0.8em} (noisy case), max sparsity $k_{\max}$.\\
\hline
1. \textbf{Initialize:}  $t=0$, $\Omega ^0=\emptyset$, $\vec{\gamma}_p^0=\vec{R}_p$, and $\rho_p^0=0$. \\
2. \textbf{While} $|\rho_p^t-\sqrt{\frac{\pi}{2}}\delta|> \theta$ and $t<k_{\max}$, \textbf{do}\\
\hspace{0.8em} a). Update the iteration index $t=t+1$. \\
\hspace{0.8em} b). Identify the support index \\\hspace{2.2em} $\varphi ^t=\arg \; \max_{j\in [1, pN]} |<\vec{\gamma}_p^{t-1},\mathbf{\Phi}_p^j>|$. \\
\hspace{0.8em} c). Update the support index set $\Omega^t=\Omega^{t-1} \cup \{\varphi ^t\}$.\\
\hspace{0.8em} d). Solve the following least squares problem and \\\hspace{2.2em} obtain a new spectral estimate: \\
\hspace{2.2em} $\hat{X}_p^t=\arg \; \min_{\vec{X}_p} \|\vec{R}_p- \mathbf{\Phi}_p(\Omega ^t) \mathbf{F}_{pN}^{-1}\vec{X}_p \|_2$. \\
\hspace{0.8em} e). Calculate the validation parameter \\\hspace{2.2em} $\rho_p^t = \frac{\|\vec{V}_p-\mathbf{\Psi}_p \mathbf{F}_{pN}^{-1} \hat{X}_p^t \|_1}{v_p}$.\\
\hspace{0.8em} f). Update the residual $\vec{\gamma}_p^t=\vec{R}_p-\mathbf{\Phi}_p \mathbf{F}_{pN}^{-1} \hat{X}_p^t$.\\
3. \textbf{Return} the spectral estimate: $\hat{X}_p= \hat{X}_p^t$. \\[1ex] 
\hline 
\textbf{Halting Criterion:} \\ \hspace{0.2em}
$\left\{ \begin{array}{ll}
\rho_p^t \le \varpi ({1-\eta}) \sqrt{\frac{2}{\pi pN}}, & \textrm{For noiseless measurements.}\\
|\rho_p^t-\sqrt{\frac{\pi}{2}}\delta|\le \theta,  & \textrm{For noisy measurements}.
\end{array} \right.$ \\
\hline\hline
\end{tabular}
\label{table:a3} 
}
\end{table}

In the SASR algorithm, the halting criterion can be adjusted when different inputs are given. For example, for noisy measurements, if the key parameter $\theta$ is of interest, we could set up $\theta$ by using an expected minimum confidence level $1-\varrho$: 
\begin{equation}
\theta=\left[ \frac{  \ln \left(\frac{2}{\varrho} \right)\delta  \pm \delta \sqrt{ \ln^2 \left(\frac{2}{\varrho}\right) + 16(4-\pi)  \ln \left(\frac{2}{\varrho}\right) v_p}}{4v_p} \right]^{+},  \label{get}
\end{equation}
where $[x]^{+}$ denotes $\max(x,0)$. We can then halt the iteration in the correct iteration index with a confidence level greater than $1-\varrho$. The proof of (\ref{get}) is similar to the proof of Theorem 2: To find the accuracy parameter $\theta$, we use the following quadratic equation regarding~$\theta$ from (\ref{rp}):
\begin{equation}
v_p \cdot \theta^2-\frac{1}{2}  \ln \left(\frac{2}{\varrho}\right) \delta  \cdot \theta-(4-\pi)\ln \left(\frac{2}{\varrho}\right) \delta^2=0.
\end{equation}
It can be easily determined that the discriminant of the above quadratic equation is positive, and we obtain the distinct real root as given by (\ref{get}). 
 
We note that one important advantage of the proposed SASR algorithm is that it does not require the knowledge of instantaneous spectral sparsity $k$; Instead, it only requires the sparsity upper bound $k_\text{max}$ which can be easily estimated by long-term spectral usage observations. Additionally, traditional greedy algorithms employ the residual $\|\gamma_p^t\|_2$ smaller than a threshold as a halting criterion, where the residual $\|\gamma_p^t\|_2$ decreases or remains as the number of iterations increases. An inappropriate threshold in greedy algorithms could lead to either under-fitting or over-fitting. By contrast, using the proposed algorithm, we monitor the validation parameter $\rho_p^t$ instead of the residual $\|\gamma_p^t\|_2$; We can terminate the iteration in the correct iteration index with a high probability which exponentially increases with $v_p$ increasing or $\delta$ decreasing. More measurements for validation can significantly reduce the risk of data under-/over-fitting. Furthermore, compared with traditional recovery algorithms, the proposed SASR algorithm reduces the number of iterations and thus the complexity. The running time of the proposed SASR algorithm is dominated by the step 2-b) as shown in Table II, whose cost is $\mathcal{O} (r_p p N)$ for one iteration. At iteration $t$, the least squares problem can be solved with marginal cost $\mathcal{O} (t \, r_p)$. As the iteration can be terminated at the correct index $t=k$ with a high probability, the total running time of the proposed SASR algorithm is thus $\mathcal{O} (k r_p p N)$. By contrast, as discussed in [33], the total running time of the traditional OMP algorithm is $\mathcal{O} (k_\text{max}M_p p N)$, as $k_\text{max}$ iterations are likely needed (i.e. an overrun occurs) when the instantaneous spectral sparsity is unknown. The computational complexity of the proposed SASR algorithm is thus lower than that of the OMP algorithm.

\section{Simulation Results}
\label{section5}

In our simulations, the wideband analog signal model in \cite{was1} was adopted; Thus, at a CR the wideband signal of interest can be written as
\begin{equation}
x(t)=\mathop{\sum}\limits_{l=1}^{N_b} \sqrt{E_l} B_l \cdot \textrm{sinc} \left(B_l(t - \alpha)\right) \cdot \cos \left(2\pi f_{l} (t - \alpha) \right) ,
\end{equation}
where $x(t)$ consists of $N_b$ non-overlapping subbands, and $E_l$, $B_l$, and $f_l$ denote the received power, the bandwidth, and the centre frequency of subband $l$ at the CR, respectively. The function sinc$(x)$ denotes the normalized sinc function, i.e., sinc$(x)=\frac{\sin (\pi x)}{\pi x}$, and $\alpha$ denotes a small random time offset. The major simulation parameters are listed in Table~III unless otherwise stated. The overall bandwidth of the signal $x(t)$ is $W$ (Hz). The frequency range of subband $l$ is $[f_l-\frac{B_l}{2}, f_l+\frac{B_l}{2}]$, where $f_l$ is randomly located within $[\frac{B_l}{2} \sim W-\frac{B_l}{2}]$. We note that in our simulations, we have the spectral occupancy $({\mathop{\sum}_{l=1}^{N_b}B_l})/{W} $ calculated as $0\%\sim 8\%$ according to the set up in Table~III, which is particularly relevant to practical CR networks. The sparsity level $k$ thus exists in the range of $0\% N\sim 8\% N$; Given a fixed value of $k$, the selection of $B_l$ will be conditional.
In addition, during the spectrum sensing duration, we assume the signal from primary users and the channel conditions are quasi-stationary. We adopt the sub-Nyquist rate $f_s$ ($f_s<2W$) for sampling the wideband signal throughout simulations and employ compressive measurement matrices with standard normal distribution. Please also note that, the size of compressive measurements is closely related to the choice of $\tau$ because $M_p=f_s p \tau$. A smaller $\tau$ will not provide a satisfactory spectral recovery rate due to insufficient training data. On the other hand, a larger $\tau$ will require more memory space to store the compressive measurements. Here, we assume $\tau=$ 0.2 $\mu$s considering both the spectral recovery requirement and memory requirement. Using the settings in  Table~III, instead of $N=2W\tau=1000$ Nyquist samples, we have $f_s \tau=200$ measurements in each time slot, among which $v_p$ measurements are used for validation and the residual is used for recovering the spectrum.

\begin{table}[ht]
\caption{Simulation Parameters for ACSS}
\begin{center}
{\small
\begin{tabular}{clr}
\toprule
\multicolumn{2}{c}{ACSS System Parameters} \\
\cmidrule(r){1-2}
Symbol & Description & Settings\\
\midrule
$W$ & Signal bandwidth of interest & 2.5 GHz \\
$N_b$ & Number of subbands & 4 \\
$k$& Spectral sparsity level & 32\\
$B_l$ &  Bandwidth of subband $l$ & 0 $\sim$ 50 MHz\\
$f_l$ & Center frequency of subband $l$ & $\frac{B_l}{2} \sim W-\frac{B_l}{2}$ \\
$\frac{E_l}{\delta^2}$ & Received SNR of subband $l$ &7 $\sim$ 25 dB\\
$\alpha$ & Small random time offset & $0\sim 0.1$ $\mu$s\\
$L$& Frame length & 4 $\mu$s\\
$T_{\min}$& Min data transmission time & 2.4 $\mu$s\\
$\tau$& Small time step & 0.2 $\mu$s\\
$f_s$& Sub-Nyquist sampling rate & 1 GHz\\
\bottomrule
\end{tabular}
}
\end{center}
\label{default}
\end{table}%

\begin{figure}[ht]
\centerline{\includegraphics[width=5.8in]{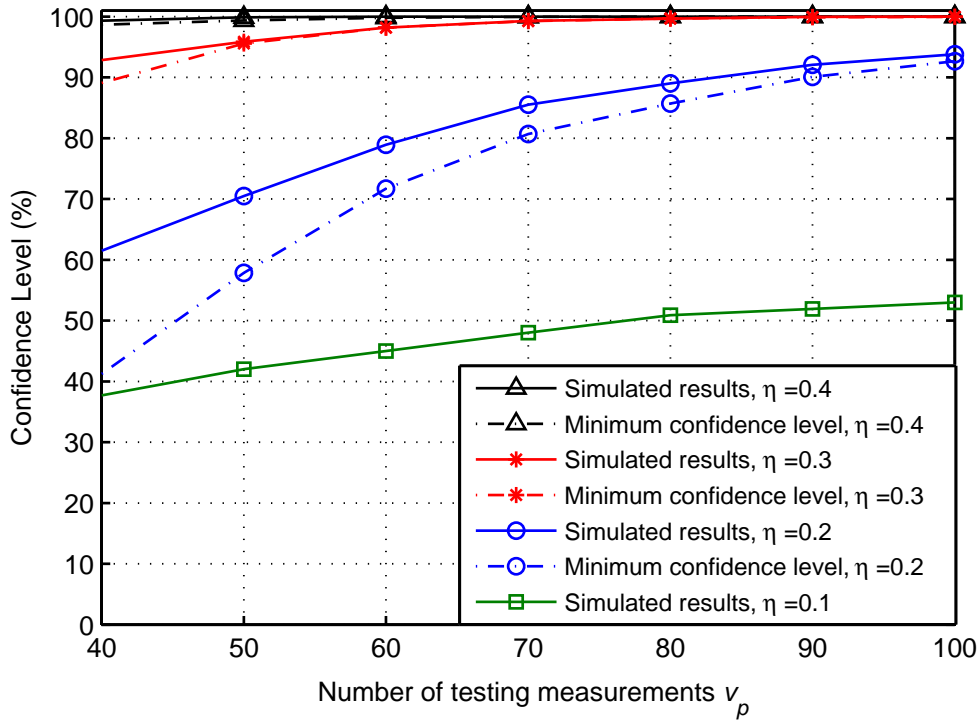}}
\vspace{-1em}
\caption{Confidence level in Lemma 1 and the effects of the confidence factor $\eta$ and the number of testing measurements $v_p$.}
\label{fig4}
\end{figure}

Firstly, in Fig.~\ref{fig4} we verify the validity and accuracy of the confidence interval shown in Lemma 1 using the settings in  Table~III. Effects of the confidence factor $\eta$ and the number of testing measurements $v_p$ on the confidence level are also demonstrated. The value of $C$ in Lemma 1 depends on the concentration property of random normal distributed variables in the matrix $\mathbf{\Psi}$, and without loss of generality we choose $C=1$ to obtain a theoretical minimum confidence level in this figure. The confidence level shown in Fig.~\ref{fig4} represents how often the actual spectral recovery error lies within the confidence interval. We can see that the wider the confidence interval we are willing to accept (with using a larger $\eta$), the more certain we can be that the actual recovery error would be within that estimated range (i.e., a higher confidence level obtained). It can also be seen that the confidence level improves with $v_p$ increasing; That is, with more testing data, validation results are more trustworthy. The minimum confidence level shown in Fig.~\ref{fig4} indicates a theoretical lower bound of how sure the estimation range can be for given settings of $\eta$ and $v_p$. With either $\eta$ or $v_p$ increasing, the lower bound is more close to the simulated confidence level. 

\begin{figure}[ht]
\centerline{\includegraphics[width=5.8in]{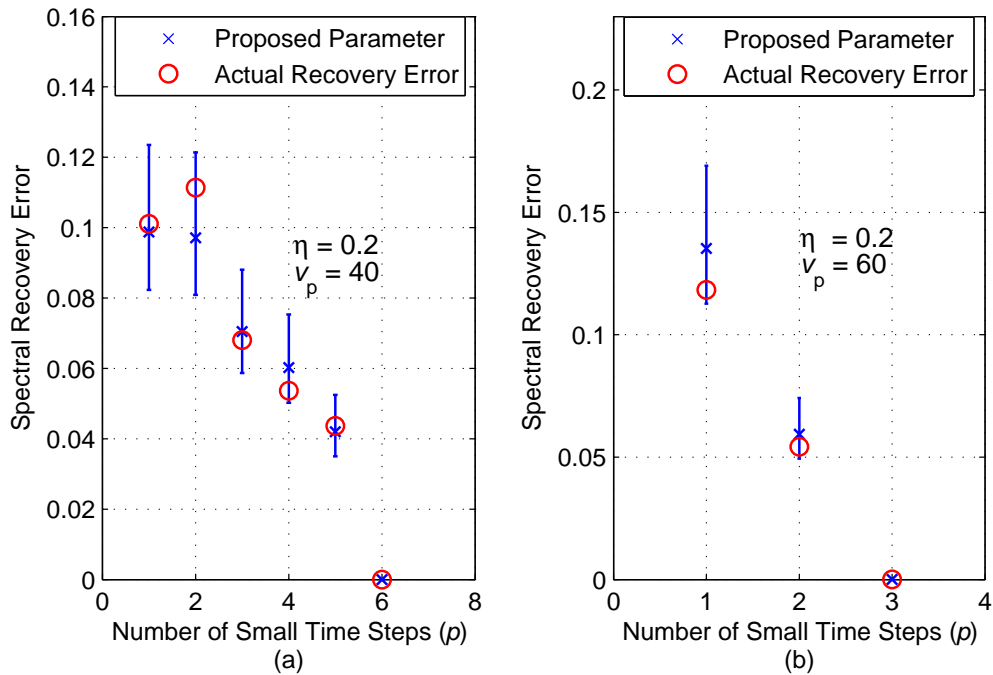}}
\vspace{-1em}
\caption{The comparison of the proposed validation parameter $\sqrt{\frac{\pi pN}{2}}\rho_p$ in Lemma 1 and the actual recovery error $\|\vec{X}_p-\hat{X}_p\|_2$ when the number of time steps increases. The blue bars give the confidence interval in Lemma 1 when the confidence factor $\eta=0.2$. (We consider the number of testing measurements $v_p=40$ in (a), and $v_p=60$ in (b).)}
\label{fig5}
\end{figure}

Using the above settings in Table III, in Fig.~\ref{fig5} we present the proposed validation parameter $\sqrt{\frac{\pi pN}{2}}\rho_p$, the actual recovery error $\|\vec{X}_p-\hat{X}_p\|_2$, and the proposed confidence interval $\left[\frac{\sqrt{\frac{\pi pN}{2}}\rho_p}{1+\eta}, \frac{\sqrt{\frac{\pi pN}{2}}\rho_p}{1-\eta}\right]$ when the number of time steps increases. To make the confidence interval narrower and more precise, we consider $\eta=0.2$, and show the effects of changing the number of testing measurements $v_p$ by using two sub-figs. We can see that the proposed validation parameter is very close to the simulated recovery error regardless the number of time steps or the value of $v_p$ varying, and can therefore be used to predict the actual recovery error. With $p$ increasing, the sensing duration is increased step by step, and the sensing will be halted if the recovery error is sufficiently small, for example we need $p=$ 6 in Fig.~\ref{fig5} (a) and $p=$ 3 Fig.~\ref{fig5} (b). It is also illustrated that the more testing measurements, the fewer time slots are required to recover the spectrum. The remaining time slots can then be used for data transmission to improve system throughput. 

\begin{figure}[ht]
\centerline{\includegraphics[width=5.8in]{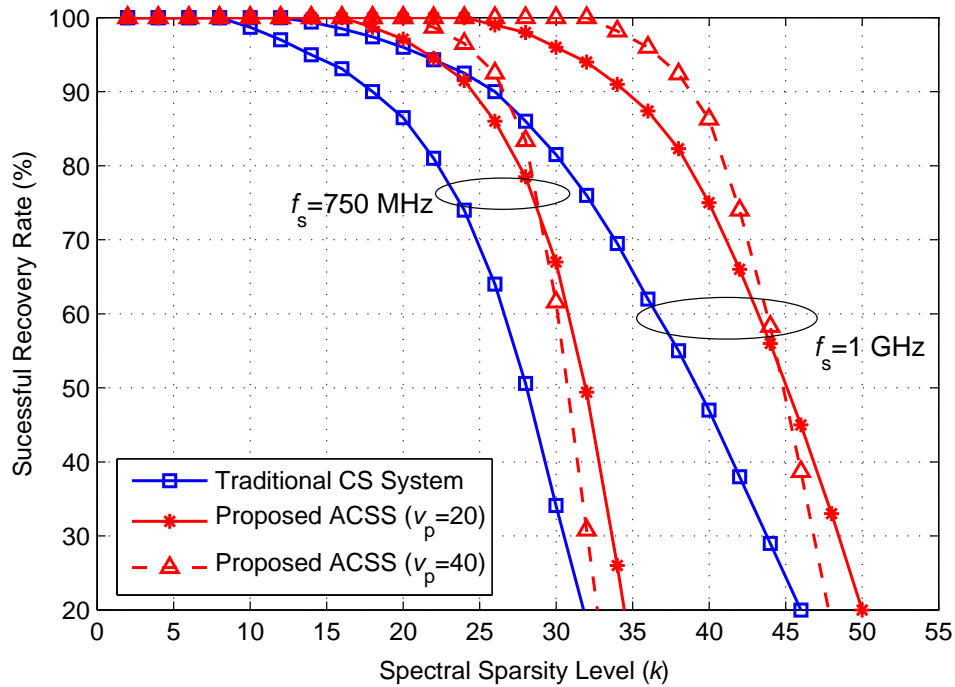}}
\vspace{-0.5em}
\caption{The performance comparison of the proposed ACSS system and the traditional CS system \cite{scs1} when the spectral sparsity level $k$ and the sub-Nyquist sampling rate $f_s$ vary. Successful spectral recovery is defined as the spectral recovery with the mean squared error not larger than 0.001.}
\label{fig6}
\end{figure}

Applying the halting criterion in Theorem 1, we now demonstrate the performance of the proposed ACSS compared to a traditional CS system in Fig.~\ref{fig6} when the spectral sparsity level $k$ varies. We consider two cases of the sub-Nyquist sampling rate, $f_s=750$ MHz and 1GHz respectively, and $\eta=0.2$. We define the successful spectral recovery as the case with the mean squared error not larger than 0.001. It is evident that the proposed ACSS can not only automatically adapt the number of measurements to the unknown sparsity level $k$, but also considerably improve the spectral recovery performance compared with the traditional CS approach no matter for either value of $f_s$.  The lower the spectral level, the higher the successful recovery rate obtained. It is also illustrated that a larger number of validation measurements $v_p$ does not always guarantee a better recovery performance: The two red curves crossover with $k$ increasing. It is because that for a fixed set of compressive measurements, a larger value of $v_p$ means a smaller training subset used for recovery which may lead to worse spectral recovery performance especially for a higher sparsity level.  

\begin{figure}[ht]
\centerline{\includegraphics[width=5.8in]{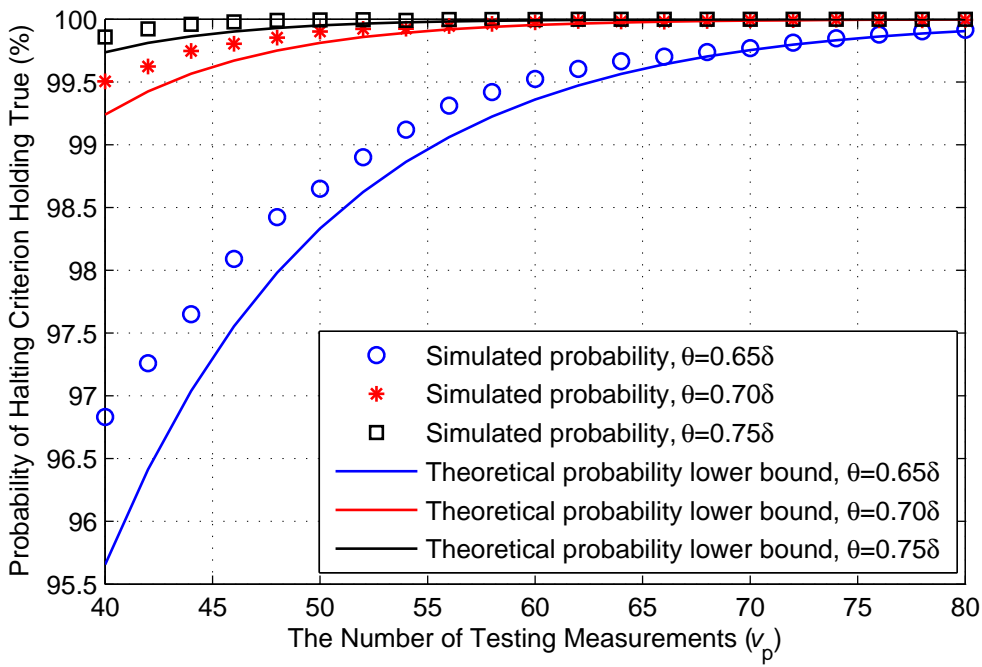}}
\vspace{-1em}
\caption{The comparison between the simulated probability of the halting criterion $|\rho_p-\sqrt{\frac{\pi}{2}}\delta|\le \theta$ holding true and the theoretical probability lower bound $1-\varrho$ in Theorem 2 when the number of testing measurements $v_p$ and the accuracy parameter $\theta$ vary.  }
\label{fig7}
\end{figure}

We now extend the use of ACSS in noisy measurement environments. Fig.~\ref{fig7} shows the comparison between the simulated probability of the halting criterion $|\rho_p-\sqrt{\frac{\pi}{2}}\delta|\le \theta$ holding true and the theoretical probability lower bound $1-\varrho$ in Theorem 2 when the number of testing measurements $v_p$ and the accuracy parameter $\theta$ vary. To guarantee a high confidence level, we consider $\theta=0.6\delta$, $0.65\delta$, and $0.7\delta$. It is shown that the lower bound is very tight and thus can be used to predict the actual probability. With a high probability of the halting criterion holding true, we can expect that a good estimation of the spectrum is found. Fig.~\ref{fig7} also shows that given a fixed confidence level of the halting criterion, at the expense of accuracy (i.e. a larger value of $\theta$), we can use fewer testing measurements. In addition, the confidence level exponentially increases with $v_p$ increasing. That is, using more testing measurements, we have a better chance of finding a good spectral estimation.

\begin{figure}[ht]
\centerline{\includegraphics[width=5.7in]{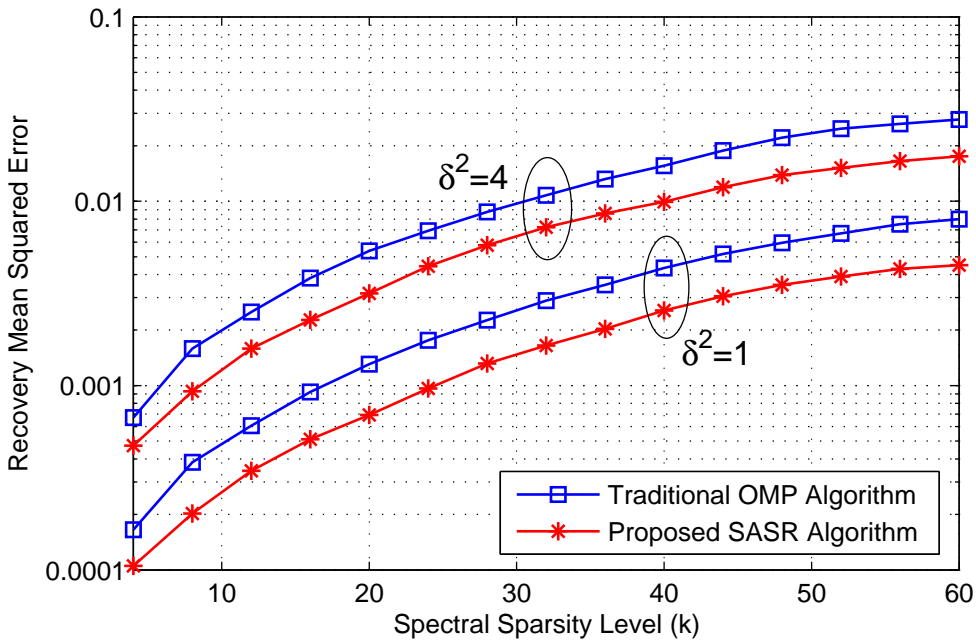}}
\vspace{-0.7em}
\caption{Performance comparison of the proposed SASR algorithm and the traditional OMP algorithm when the spectral sparsity level and the noise variance $\delta^2$ vary in noisy environments. The recovery mean squared error is defined as $\mathbb{E} \left[(\vec{X}_{p,i}-\hat{X}_{p,i})^2/\vec{X}_{p,i}^2 \right]$ where $\vec{X}_{p,i}$ denotes the $i$-th component of the vector $\vec{X}_{p}$.}
\label{fig8}
\end{figure}

Fig.~\ref{fig8} shows the performance comparison of the proposed SASR algorithm and the traditional OMP algorithm when the actual spectral sparsity level $k$ and the noise variance $\delta^2$ vary in noisy environments. We assume the sparsity level $k$ is unknown when performing recovery, but we know that $k$ exists in the range of $0\% N\sim 8\% N$, i.e., $k_{\text{max}}=8\% N=80$ according to the settings in Table III. The received signal-to-measurement-noise (SNR) ratios of these subbands are set to be randomly distributed between $7\sim 25$ dB as listed in Table III. We consider $\delta^2=1$ and 4, respectively, and $v_p=$40.  The recovery mean squared error in the noisy case is defined as $\mathbb{E} \left[(\vec{X}_{p,i}-\hat{X}_{p,i})^2/\vec{X}_{p,i}^2 \right]$ where $\vec{X}_{p,i}$ denotes the $i$-th component of the vector $\vec{X}_{p}$. We can see that compared to the traditional OMP algorithm, the proposed SASR provides much better spectral estimation and recovery performance, regardless the values of $\delta^2$ or $k$. It is because that the OMP algorithm tends to use more number of iterations to avoid under-fitting problems and to prevent missed detection leading to harmful interference to PUs in CR networks. However, on the other hand, using more number of iterations will cause over-fitting problems and exaggerate minor fluctuations in the data which will finally result in poor recovery performance. We would like to emphasize that the proposed SASR algorithm will obtain a more significant performance improvement in practice, as there always exists a larger uncertainty of $k$ in realistic wideband CR networks.

\section{Conclusions}
\label{section6}

In this paper, we have proposed a novel framework, i.e. ACSS, for compressive spectrum sensing in wideband CR networks. ACSS enables a CR to automatically adopt an appropriate number of compressive measurements without knowledge of the instantaneous spectral sparsity level, while guaranteeing the wideband spectrum recovery with a small predictable recovery error. This is realized by the proposed measurement procedure and the validation approach. The validation approach can accurately estimate the actual spectral recovery error with high confidence by using only a small amount of testing data. The proposed ACSS thus avoids excessive or insufficient numbers of compressive measurements, and helps enhance the recovery performance and improve the energy efficiency of CR networks. In addition, we extend the use of ACSS to noisy environments and propose another validation approach: If a good spectral estimate exists, the validation approach will find it with a high probability. Furthermore, we have proposed the SASR algorithm to recover the wideband spectrum without requiring the knowledge of the instantaneous spectral sparsity level. The SASR algorithm can autonomously adopt a proper number of iterations, and thus  solve the under-fitting or over-fitting problems which commonly exist in most other greedy recovery algorithms.

Simulation results have shown that the proposed ACSS framework can correctly stop the signal acquisition that saves both spectrum sensing time and signal acquisition energy in both noiseless and noisy environments. Compared to traditional CS, ACSS can not only provide better spectral recovery performance, but also help improve system throughput and energy efficiency of CR networks. In addition, the proposed SASR algorithm can achieve lower recovery mean squared error and better spectrum sensing performance compared to the OMP algorithm.  
We emphasize that the ACSS framework is not limited to CR networks; The proposed validation approach could be extended to other CS applications, e.g., a CS enabled communication system where the approach could be used to terminate signal detection at an appropriate time.  
Since RF spectrum is essential to wireless communications and the wideband techniques could potentially provide higher capacity, the proposed framework in this paper is thus particularly valuable and can have a wide range of applications, e.g., in broadband spectral analyzers and ultra wideband radars.

\appendices

\section{Proof of Lemma~1}
The Johnson-Lindenstrauss Lemma\cite{JL} states that a set of $N$ points in a high-dimensional Euclidean space can be mapped (with low distortion) into a Euclidean space of much lower dimension  $v_p$, and all distance are preserved up to a multiplicative confidence factor between $1-\eta$ and $1+\eta$. With the aid of the Johnson-Lindenstrauss Lemma in Theorem~5.1 of \cite{JL}, we get $v_p=C\eta^{-2}\log \frac{4}{\xi}$ where $C$ denotes a positive constant, and  
\begin{equation}
\Pr \left[ (1\!-\!\eta)\|\vec{x}\|_2 \! \le \!\frac{ \| \mathbf{\Psi}_p \vec{x} \|_1}{\sqrt{2/\pi}v_p} \le (1\!+\!\eta)\|\vec{x}\|_2 \right]\ge 1\!- \xi.
\label{proof}
\end{equation}
Replacing $\vec{x}$ in (\ref{proof}) by $\mathbf{F}_{pN}^{-1} (\vec{X}_p-\hat{X}_p)$, we have the inequality (\ref{proof1}). Jointly using (\ref{test}) and (\ref{valid}), we simplify (\ref{proof1}) to (\ref{proof2}). Applying Parseval's relation~to (\ref{proof2}), we then get (\ref{proof3}). The equations (\ref{proof1}-\ref{proof3}) are shown on the top of the next page.
\begin{figure*}[!t]
\begin{equation}
\Pr \! \left[ \!(1\!-\!\eta)\|\mathbf{F}_{pN}^{-1} (\vec{X}_p\!-\!\hat{X}_p)\|_2 \! \le \!\frac{ \| \mathbf{\Psi}_p \mathbf{F}_{pN}^{-1} (\vec{X}_p\!-\!\hat{X}_p) \|_1}{\sqrt{2/\pi}\;v_p} \! \le \!(1\!+\!\eta)\|\mathbf{F}_{pN}^{-1} (\vec{X}_p\!-\!\hat{X}_p)\|_2 \! \right] \!\ge 1\!- \xi.
\label{proof1}
\end{equation}
\begin{equation}
\Pr \left[ (1\!-\!\eta)\|\mathbf{F}_{pN}^{-1} (\vec{X}_p\!-\!\hat{X}_p)\|_2 \! \le \sqrt{\frac{\pi}{2}} \rho_p \le (1\!+\!\eta)\|\mathbf{F}_{pN}^{-1} (\vec{X}_p\!-\!\hat{X}_p)\|_2 \right]\ge 1\!- \xi.
\label{proof2}
\end{equation}
\begin{equation}
\Pr \left[(1-\eta)\|\vec{X}_p-\hat{X}_p\|_2 \le \sqrt{\frac{\pi pN}{2}}\rho_p \le (1+\eta)\|\vec{X}_p-\hat{X}_p\|_2 \right] \ge 1-\xi.
\label{proof3}
\end{equation}
\end{figure*}
And finally, we obtain
\begin{equation} 
\Pr \left[\frac{\sqrt{\frac{\pi pN}{2}}\rho_p}{1+\eta} \le \|\vec{X}_p-\hat{X}_p\|_2 \le \frac{\sqrt{\frac{\pi pN}{2}}\rho_p}{1-\eta} \right] \ge 1-\xi. 
\end{equation}
This completes the proof.  

\section{Proof of Theorem~1}

Using the probabilistic inequality $ \Pr(B)\ge \Pr(A\cap B)$, we can obtain the following inequality: 
\begin{equation}
\begin{array}{l}
\Pr \left[ \|\vec{X}_p-\hat{X}_p\|_2 \le \frac{\sqrt{\frac{\pi pN}{2}}\rho_p}{1-\eta} \right] \ge \\
\Pr \left[\frac{\sqrt{\frac{\pi pN}{2}}\rho_p}{1+\eta} \le \|\vec{X}_p-\hat{X}_p\|_2 \le \frac{\sqrt{\frac{\pi pN}{2}}\rho_p}{1-\eta} \right]. \label{good1} 
\end{array}
\end{equation}
If the halting criterion $\rho_p \le \varpi ({1-\eta}) \sqrt{\frac{2}{\pi pN}}$ is met, we have $\frac{\sqrt{\frac{\pi pN}{2}}\rho_p}{1-\eta} \le \varpi$, then the following inequality holds:
\begin{equation}
 \Pr \left[ \|\vec{X}_p-\hat{X}_p\|_2 \le \varpi \right]
 \ge 
  \Pr \left[ \|\vec{X}_p-\hat{X}_p\|_2 \le \frac{\sqrt{\frac{\pi pN}{2}}\rho_p}{1-\eta} \right]. \label{good2}  
\end{equation}
With the aid of Lemma 1, jointly using (\ref{the1}), (\ref{good1}), and (\ref{good2}), we have 
\begin{equation}
\Pr \left[ \|\vec{X}_p-\hat{X}_p\|_2 \le \varpi \right] \ge  1-\xi = 1-4 \exp(-\frac{v_p\eta^2}{C}).
\end{equation}
This completes the proof.

\section{Proof of Theorem~2}

Suppose that $\hat{X}_p$ is a good spectral estimate such that $\hat{X}_p=\vec{X}_p$, we can write the validation parameter by using (\ref{valid}) and (\ref{test1})
\begin{equation}
\rho_p = \frac{\|\vec{V}_p - \mathbf{\Psi}_p \mathbf{F}_{pN}^{-1} \hat{X}_p \|_1}{v_p} = \frac{\| \vec{n}_V \|_1}{v_p} = \frac{\sum_{i=1}^{v_p}|n_V^i |}{v_p}.
\label{valid1}
\end{equation}

Define a new variable $D_i=|n_V^i|-\sqrt{\frac{\pi}{2}}\delta$. Since the measurement noise $n_V^i \sim \mathcal{CN}(0, \delta^2)$, we have $D_i$ following the Rayleigh distribution with zero mean and variance $\frac{4-\pi}{2}\delta^2$. Additionally, 
we can find that $|D_i|\le 3\delta$ with $99.7\%$ confidence (according to the three-sigma rule) which is like being almost sure.
Using the Bernstein's inequality \cite{bein}, we obtain the following inequality:
\begin{equation}
\begin{array}{l}
\Pr \left[ \left|\sum_{i=1}^{v_p} D_i \right|> \zeta \right] = \Pr \left[ \left|\sum_{i=1}^{v_p} |n_V^i| - v_p \sqrt{\frac{\pi}{2}}\delta \right|> \zeta \right] \\
 \le 2\exp \left( -\dfrac{\zeta^2/2}{ \sum_{i=1}^{v_p} \mathbb{E}[ D_i^2 ] + \max(|D_i|) \zeta/3}\right) \\
=  2\exp \left( -\dfrac{\zeta^2}{(4-\pi)v_p\delta^2 + 2\zeta \delta }\right).
\label{bba}
\end{array}
\end{equation}
Letting $\zeta=v_p\theta$ and using (\ref{valid1}), we can rewrite the above inequality
\begin{equation}
\Pr \left[ \left| \rho_p - \sqrt{\frac{\pi}{2}}\delta \right| > \theta \right] \le 2\exp \left(-\frac{v_p\theta^2}{(4-\pi)\delta^2+2 \theta \delta } \right). \label{xiao}
\end{equation}
Equivalently, (\ref{xiao}) can be written as
\begin{equation}
\Pr \left[ \left| \rho_p - \sqrt{\frac{\pi}{2}}\delta \right| \le \theta \right] > 1- 2\exp \left(-\frac{v_p\theta^2}{(4-\pi)\delta^2+2  \theta \delta } \right). \label{prob}
\end{equation}
Aligning the right item of (\ref{prob}) with the lower bound $1-\varrho$, after manipulation we obtain
\begin{equation}
v_p=\ln \left(\frac{2}{\varrho}\right) \frac{(4-\pi)\delta^2+2 \theta \delta }{\theta^2}.
\label{rp}
\end{equation}
This completes the proof of Theorem~2.



\bibliographystyle{IEEEtran}
\bibliography{cwpf}

\begin{IEEEbiography}[{\includegraphics[width=1in,height=1.25in,clip,keepaspectratio]{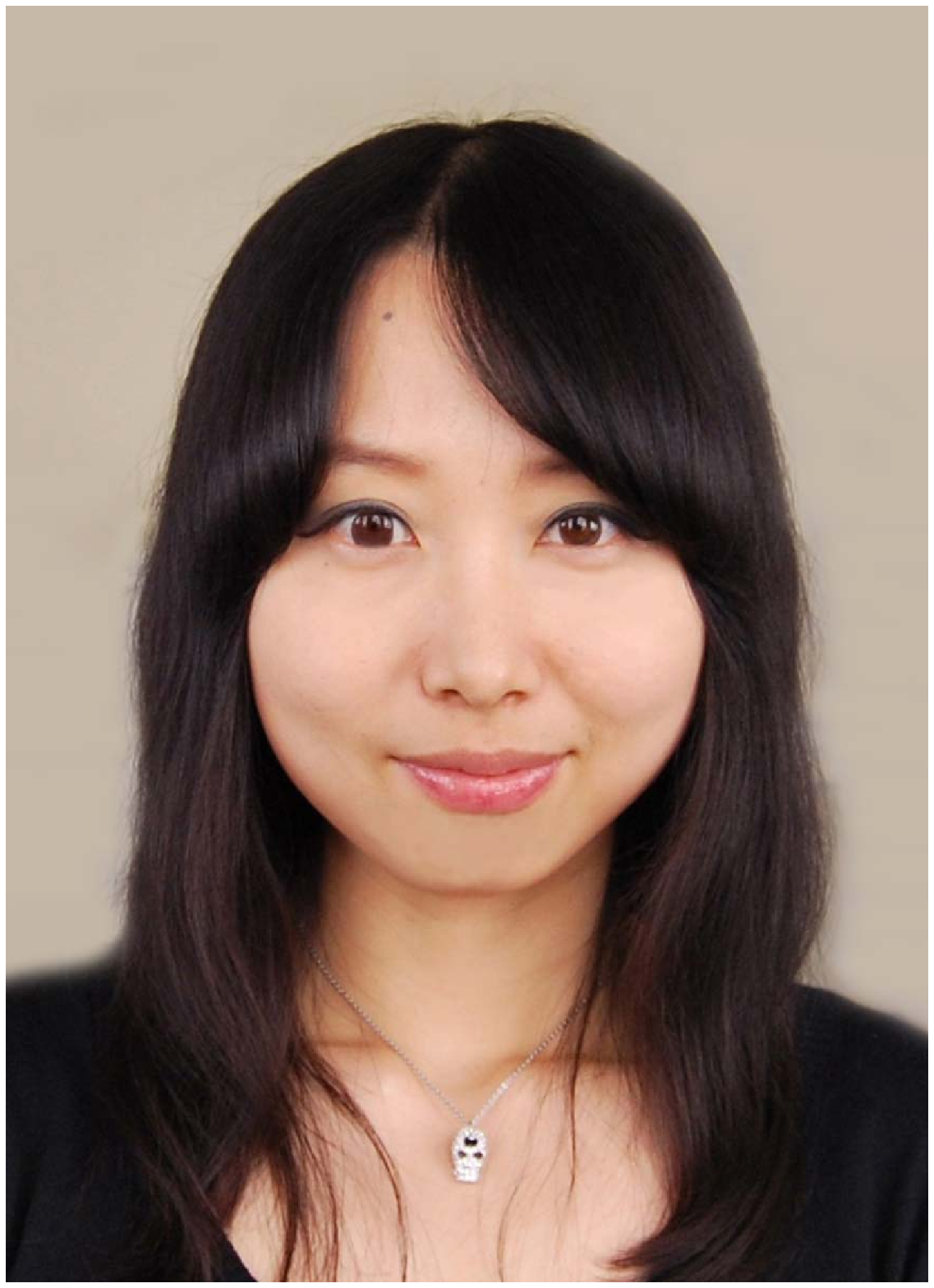}}]
{Dr Jing Jiang} (S'10-M'13) received her B.Eng. and M.Sc. degrees from Harbin Institute of Technology, China, in 2005 and 2007, respectively, and the Ph.D. degree in electronic engineering from the University of Edinburgh, UK, in 2011. She then joined University of Surrey, UK, as a postdoctoral research associate in 2011.
She is now a research associate with the Institute for Automotive and Manufacturing Advanced Practice (AMAP) at the University of Sunderland, UK. Her recent research interests include 5G wireless communications, massive-MIMO systems, MIMO and virtual-MIMO systems, cognitive radio systems, energy-efficient system design, compressive sensing techniques, relay and cooperation techniques, energy saving techniques for electric vehicles, digital technologies in advanced manufacturing, and information and communications technologies in green vehicles.
\end{IEEEbiography}
\smallskip{}
\begin{IEEEbiography}[{\includegraphics[width=1in,height=1.25in,clip,keepaspectratio]{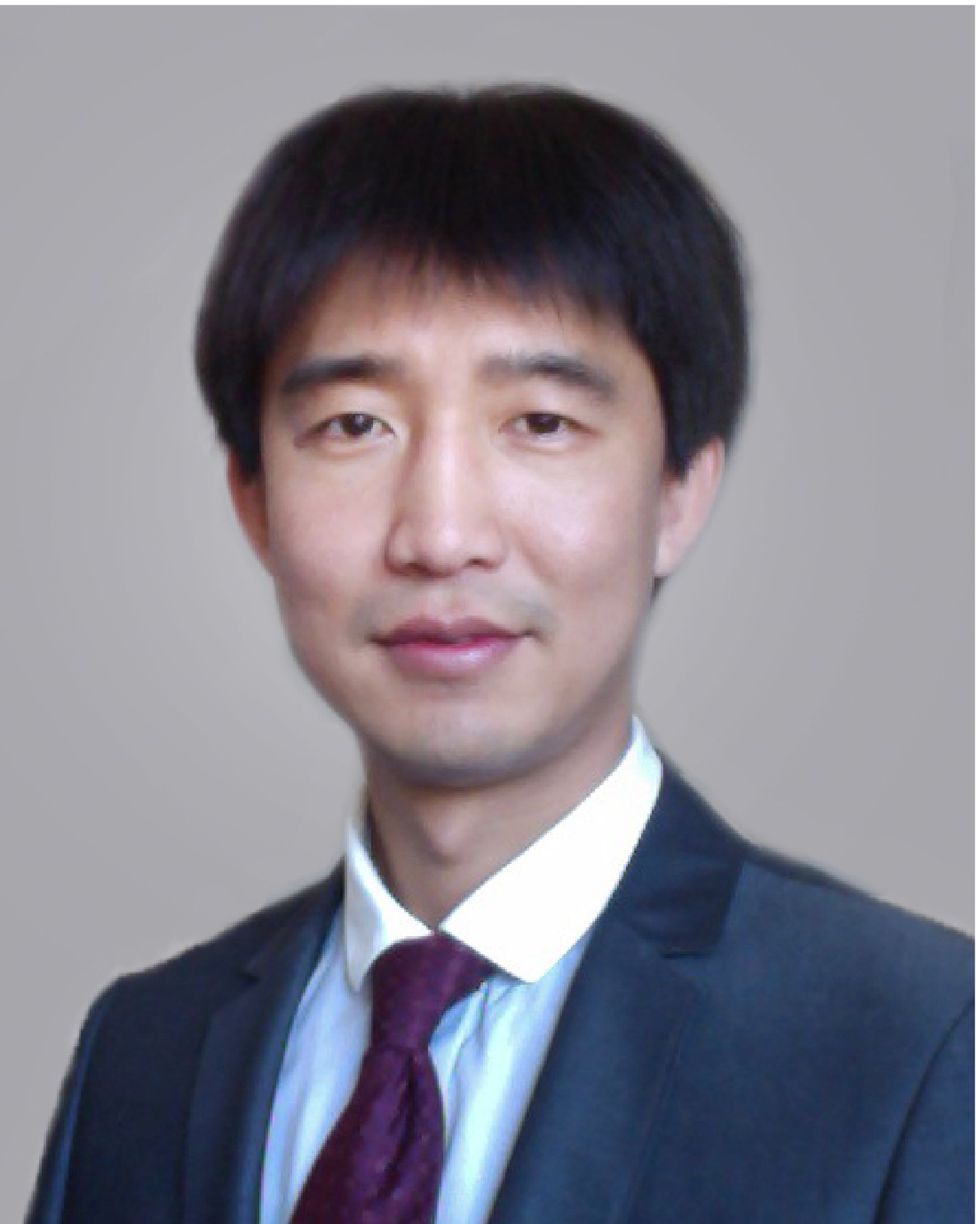}}]
{Dr Hongjian Sun} (S'07-M'11-SM'15) received his Ph.D. degree in 2010 at the University of Edinburgh, UK. He then joined King's College London, UK, as a Postdoctoral Research Associate in 2010. In 2011-2012, he was a visiting Postdoctoral Research Associate at Princeton University, USA. Since 2013, he has been a Lecturer in Smart Grid at the University of Durham, UK. His recent research interests include Smart Grids, Wireless Communications, and Signal Processing. He has made 1 contribution to the IEEE 1900.6a Standard, and published 2 book chapters and more than 50 papers in refereed journals and international conferences. 

He is on the Editorial Board for Journal of Communications and Networks, and EURASIP Journal on Wireless Communications and Networking, and was a Guest Editor for the special issue ``Industrial Wireless Sensor Networks" for International Journal of Distributed Sensor Networks. Additionally, he is serving as an organizing chair for Workshop on Integrating Communications, Control, Computing Technologies for Smart Grid, Glasgow, UK, in May 2015, and Workshop on Communications Technologies for Smart Grid, Shanghai, China, in August 2015. He also served (or is serving) as a technical program committee (TPC) member for many international conferences, e.g., ICC, Globecom, VTC. He is a peer-reviewer for a number of international journals and was nominated as an Exemplary Reviewer by IEEE Communications Letters in both 2011 and 2012.  
\end{IEEEbiography}
\smallskip{}
\begin{IEEEbiography}[{\includegraphics[width=1in,height=1.25in,clip,keepaspectratio]{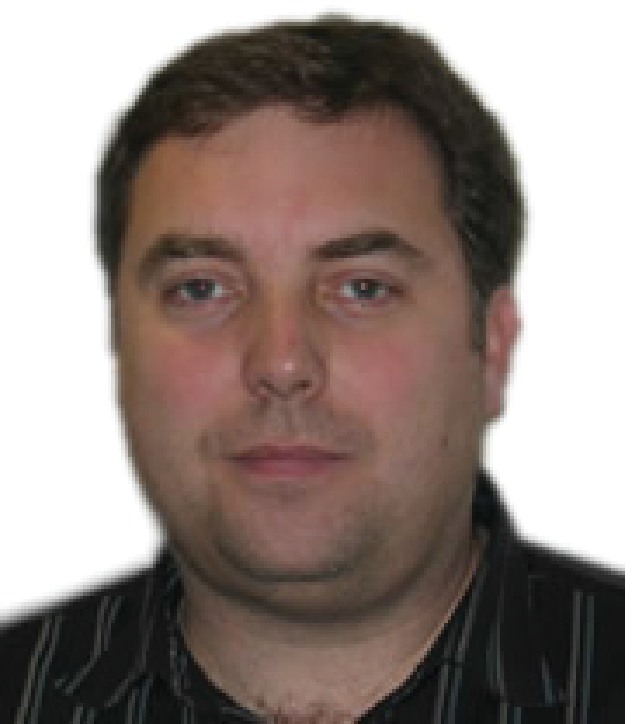}}]
{Dr David Baglee} gained his PhD from the University of Sunderland in 2005.  He is a Senior Lecturer and Project Manager at the University of Sunderland, UK and a Visiting Professor in Operations and Maintenance at the University of Lulea, Sweden and a Visiting associate Research Professor at the University of Maryland USA. His research interests include the use of advanced maintenance techniques and technologies to support advanced manufacturing within a range of industries and maintenance within ultra low carbon technologies including wind turbines, electric vehicles and hydrogen fuel cells. He has published extensively in international journals and attended a large number of international conferences.  He has managed a number of European funded projects working with BP, Nissan, Fiat and Volvo, and is a member of Euronseam, a group of academic and industrial specialists in maintenance.
\end{IEEEbiography}
\smallskip{}
\begin{IEEEbiography}[{\includegraphics[width=1in,height=1.25in,clip,keepaspectratio]{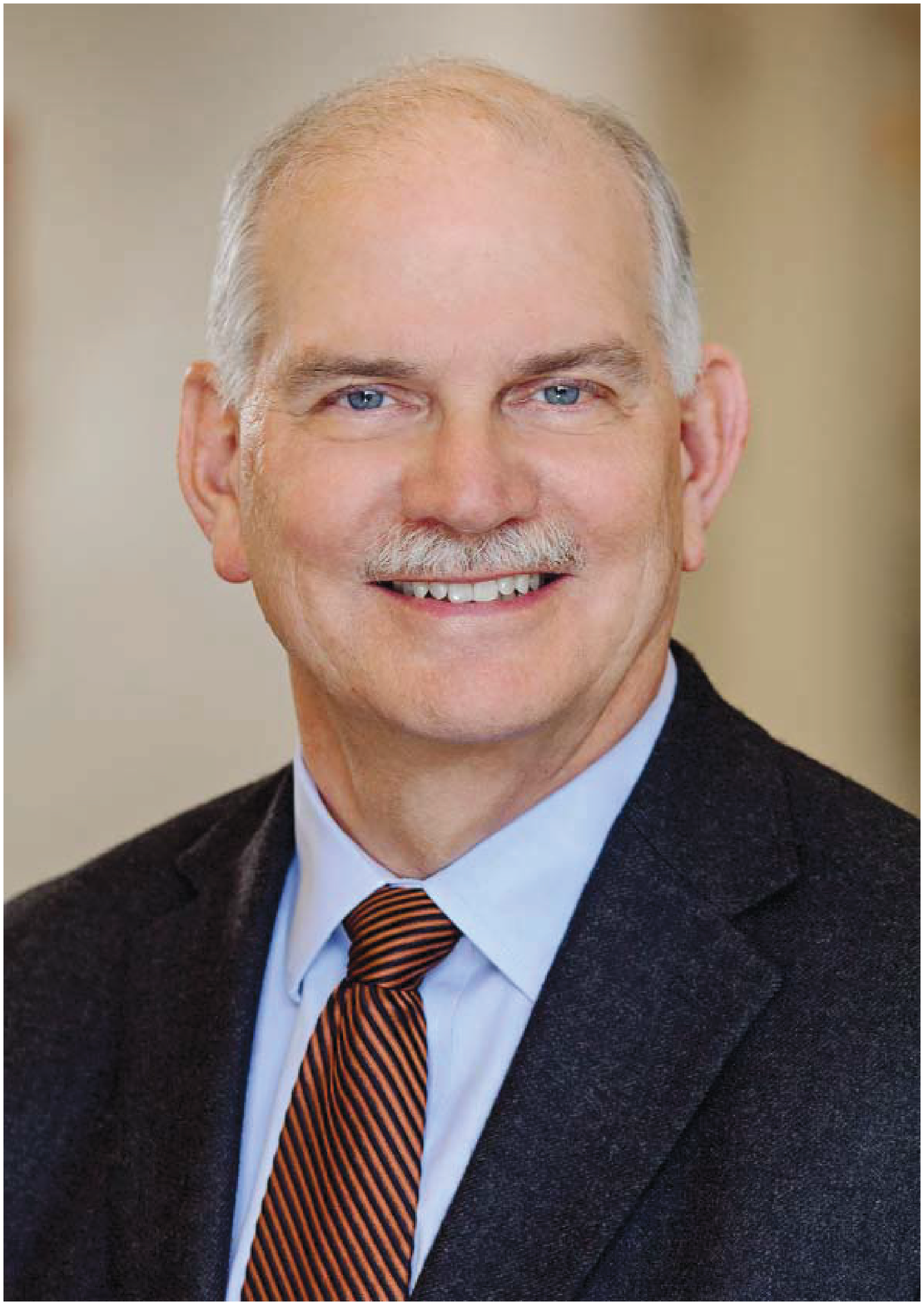}}]
{Professor H. Vincent Poor} (S'72, M'77, SM'82, F'87) received the Ph.D. degree in EECS from Princeton University in 1977.  From 1977 until 1990, he was on the faculty of the University of Illinois at Urbana-Champaign. Since 1990 he has been on the faculty at Princeton, where he is the Michael Henry Strater University Professor of Electrical Engineering and Dean of the School of Engineering and Applied Science. Dr. Poor’s research interests are in the areas of stochastic analysis, statistical signal processing, and information theory, and their applications in wireless networks and related fields. Among his publications in these areas is the recent book Mechanisms and Games for Dynamic Spectrum Allocation (Cambridge University Press, 2014).

Dr. Poor is a member of the National Academy of Engineering and the National Academy of Sciences, and a foreign member of Academia Europaea and the Royal Society. He is also a fellow of the American Academy of Arts and Sciences, the Royal Academy of Engineering (U. K), and the Royal Society of Edinburgh. In 1990, he served as President of the IEEE Information Theory Society, and in 2004-07 he served as the Editor-in-Chief of the IEEE Transactions on Information Theory. He received a Guggenheim Fellowship in 2002 and the IEEE Education Medal in 2005. Recent recognition of his work includes the 2014 URSI Booker Gold Medal, and honorary doctorates from several universities, including Aalto University in 2014.
\end{IEEEbiography}

\vfill

\end{document}